# Preferential attachment: a multi-attribute growth process generating scale-free networks of different topologies

**Dimitrios Tsiotas**
Department of Planning and Regional Development, University of Thessaly,
Pedion Areos, Volos, 38334, Greece
Tel +302421074446, fax: +302421074493
E-mail: tsiotas@uth.gr

**Abstract**

This paper expands the degree-based consideration of the preferential attachment growth process and applies five different connectivity criteria (node degree, clustering coefficient, betweenness centrality, closeness centrality, and eigenvector centrality) to define the development of new links in the networks. Based on statistical inference, the analysis shows that all the available control attributes are capable generating SF networks, that the proposed generalized preferential attachment growth process produces networks of statistically different topologies, under different control-attributes, and that the betweenness centrality is the control-attribute generating networks of better topology. Overall, this paper introduces a multi-dimensional conceptualization of preferential attachment, which can motivate further research and can provide new tools for the modeling and interpretation of real-world networks that currently cannot be fully explained by the degree-driven BA models.

## 1. Introduction

The scale-free (SF) property has become a major concept in the study of complex networks [1,R2] because empirical research has shown that many real-world networks, such as biological, citation, spatial, economic, technological, and social networks [2,4-6] are SF [7]. Generally, a network is considered as SF when its degree distribution $p(k)$ follows asymptotically a power-law (PL) pattern, of the form [1]:

$$p(k) \sim k^{-\gamma} \qquad (1),$$

where $k$ is node degree and $\gamma$ is the PL exponent, which needs to be $\gamma > 1$ in order the Riemann zeta function to be finite [3]. Research in real-world networks has shown that SF networks usually have their PL exponent ($\gamma$) ranging within the interval $2<\gamma<3$ [1], although this is not a defining condition and it may exceed these "typical" bounds [8].

In network science, the most common method for generating SF networks was proposed by Barabasi and Albert [2] and it is known as the *Barabasi-Albert* (*BA*) model. The generative mechanism of the *BA* model is based on growth and on the preferential attachment (PA) process [1,9], according to which SF networks are produced over time when the probability for new connections is proportional to node degrees. In particular, in the PA process, new nodes that are entered in the network "prefer" (wherefrom the term preferential comes from) to be connected with the already highly connected nodes (the so-called hubs) in order to benefit of the connectivity advantage of the latter. This procedure leads to the emergence of hierarchies, where hubs undertake the major load of connectivity in the network and they preserve it at the future growth of the network, a fact that is reflected on the PL shape of the degree distribution curve [7].

The PA process is very important in network science because it generates *BA* networks, which abound in the scientific literature and they have thus become the standard SF reference model in the related research [7]. However, PA does not originate from the research field of complex networks, but it was the result of the multidisciplinary nature of network science. Substantially, the PA process originates from the stochastic "Yule process" introduced by the British statistician George Udny Yule [10] during the study of evolution of species. Further, the rationale behind PA can be found almost a century ago, in sociology, in the so-called "Matthew effect" [11], which is summarized by the motto "the rich get richer and the poor get poorer", or even in economics, in the Gibrat's law [12], which describes the proportional growth of firms in terms of their absolute size.

Despite the multidisciplinary rationale observed in the conceptual framework of the PA process, in network science, this mechanism generating SF networks is limited and has one-dimensional use, since it is based only on degree. That is, the connectivity criterion (i.e. the criterion of creating new links) that has been diachronically applied in the PA process to generate SF networks was exclusively node-degree [1,4,6,7], which defines the probability of new connections added in the network to be proportional to the degree of the target-nodes [7]. Although this degree-based criterion was proven fruitful in the evolution of network science [1,4,6,7], it suggests a restriction in the PA process, whether taking into account the variety of network attributes or measures that can be used as attractors defining the PA's connectivity. Namely, instead of using node-degree, connectivity in the PA process can excellently be defined by other determinants (node-attributes), such as the clustering coefficient [4,13], the centrality measures [14], etc.

The only exemption that can be found in the degree-based consideration of the PA process is the recent work of [15], the authors of which pioneered in considering the weighted betweenness centrality as a new determinant of connectivity in the PA process, instead of the by default used degree. Focusing on how social network dynamics can be better explained, they observed that the connectivity criterion of degree, in the PA process, is not the main attractor of new social links and thus that the degree-driven PA cannot fully explain the social network dynamics. This made them consider betweenness centrality as an alternative determinant of connectivity in the PA process and led them introducing the weighted betweenness PA (WBPA) model, which reproduced more accurately a wide range of real-world social networks. The authors concluded that node-betweenness suggests a better indicator of social attractiveness and they interpreted, from a socio-psychological point of view, that the betweenness attractor, in the WBPA model, impels "individuals to (intuitively) perceive node's betweenness as the capacity of bridging communities, irrespective of its degree".

Being motivated by this insightful work, this paper extends the novel approach of [15], to consider betweenness as a new determinant of connectivity in the PA process, and it generalizes the conceptualization of PA into a multi-attribute growth process generating networks under different connectivity criteria defined by various node-attributes in a network. In particular, this paper applies five different determinants of connectivity (henceforth will be called as control-attributes) in the PA process, which are defined by the node-measures of degree, clustering, betweenness, closeness, and eigenvector centrality, and it examines the differences between the network topologies produced by each process. The research questions that are examined in this paper are, first, whether the PA process can generate SF networks for every control-attribute (determinant of connectivity), next, whether the network topologies produced by the generalized PA (GPA) process are different, and, finally, which control-attribute is capable producing a better network topology. The analysis aiming to answer these questions is based on simulations, where the networks generated by the GPA process under different control-attributes are being statistically tested and compared. By answering these questions, this paper contributes to scientific research by introducing a multi-dimensional (GPA) conceptualization of the PA process, which can motivate further research on pattern recognition between SF networks [7], and it can provide new tools for modeling and interpreting real-world networks that currently cannot be fully explained by the degree-driven BA model, as the authors of [15] note.

The remainder of this paper is organized as follows; Section 2 describes the GPA models' construction and simulation, Section 3 shows the simulation results and the statistical analysis applied to the topological attributes of the available models produced by the GPA process, under different control-attributes. Finally, in Section 4 conclusions are given.

## 2. Null models' construction

For the simulations, undirected null models are generated based on the rationale of the uniform attachment algorithm of the *BA* model [2]. An implementation of this algorithm is available in the open-source software of [16] (version 0.8.2), where its initial parameters are the number of nodes ($n$) in the generated network, the number of nodes at the start time ($m_o$), and the number of edges coming with every new node ($m$). This algorithm is subjected to the following restrictions [16]:

$$\begin{cases} n > 0 \\ 0 < m_o < n \\ 0 < m \leq m_o \end{cases} \quad (1).$$

The *BA* model's algorithm (uniform attachment) was customized (expanded) in order to be capable implementing the GPA process under different control attributes. The expansion of this algorithm (*BAmXA*) was written in Matlab code (*m*-file) [17]. In *BAmXA*, the parameter $m_o$ is set by default to one ($m_o=1$) and a new input-argument ($c$) expressing the control-attribute of GPA process' connectivity is entered in the algorithm, where values refer to the control-attribute of *degree* ($c=1$), *clustering coefficient* ($c=2$), *betweenness centrality* ($c=3$), *closeness centrality* ($c=4$), and *eigenvector centrality* ($c=5$) (for descriptions of these measures see [4,7,13]), respectively. At the first step ($p=1$) of *BAmXA*, all $n$ in number nodes have uniform probability (initial attractiveness, see [18]) to develop a connection, which equals to $P_o(i)=1/n$, where $i$ expresses a node. Based on the initial state of attractiveness ($P_o$), a pair of nodes ($i,j$) is randomly being chosen to develop a connection. At every next step ($p:=p+1$) of the algorithm, a node's $i$ initial attractiveness $P_o(i)$ is augmented with a preferential probability $P_p(i)$, defined by the value $c_p(i)$ of the control-attribute ($c=1,…,5$) for the certain node ($i$), as this value $c_p(i)$ is configured at the current stage ($p$) of the network growth. At this stage ($p$), a connection is developed proportionally-randomly to the quantities $P_p'(i)$ defined by the sums $P_o+P_p(i)$, which are normalized so that the total probability equals to one, according to the relation:

$$P_p'(i) = \frac{P_o + P_p(i)}{\sum_n (P_o + P_p(i))} \quad (2),$$

Null models generated by the *BAmXA* have size (number of nodes) ranging between 50 and 1500 nodes (see Appendix), where successive cases differ in 50 nodes. Null models bigger than 1500 nodes were not generated, due to computational constraints, and particularly because the computation of betweenness and closeness centrality within the algorithm's loops was a very time-demanding process. The reduction of the complexity of *BAmXA* and the implementation of this analysis to bigger, in size, networks suggest avenues for further research.

Due to the probabilistic structure of the algorithm, the generated null models include isolated nodes, which were ignored in the application of PL-fittings to the degree distributions of the null models.

## 3. Simulations and analysis

Simulations were conducted on 150 undirected null models, which belong to 5 families of 30 networks each (see Appendix). The $G(k)$ family includes *BA* models generated under the control-attribute of *degree* ($k$), whereas the $G(C)$, $G(CB)$, $G(CC)$, $G(CE)$ families include null models of unmapped topologies, generated under the control-attributes of *clustering coefficient* ($C$), *betweenness centrality* ($CB$), *closeness centrality* ($CC$), and *eigenvector centrality* ($CE$), respectively. Null models included in each family have sequentially $n=50, 100, 150, 200, \ldots, 1450, 1500$ number of nodes, implying that ordered pentads (fives) among families $\{G_i(k), G_i(C), G_i(CB), G_i(CC), G_i(CE) \mid i=1,2,\ldots,30\}$ are equivalent in network size but not necessarily in graph density, because the number of edges ($m$) of the null models within each pentad may differ due to the probabilistic architecture of the algorithm.

Network topologies of the available null models are embedded in the 2d-Euclidean space and they are visualized using the "Force-Atlas" layout which is available in the open-source software of [16]. This layout is generated by a force-directed algorithm (see [19]), which is used in its default parameters. This algorithm applies repulsion strengths between network hubs while arranging the hubs' connections into surrounding clusters.

### 3.1. Examination of the SF property

This part of analysis examines the first research question about detecting whether the null models produced by the GPA process have the SF property. This is by default expected only for the $G(k)$ family and it is under evaluation for the others. At first, the degree distributions of the available families of null models are illustrated into 3d bar-charts (Fig.1), where the *x*-axis represents the degree classes, the *y*-axis represents the null models' ranking (in ascending order) according to their number of nodes (i.e. 1:=50nodes, 2:=100nodes, 3:=150nodes , …, 29:=1450nodes, 30:=1500nodes), and the *z*-axis represents node-frequencies (i.e. the number of nodes lying under a certain degree). Axes have fixed minimum and maximum values so that to be comparable. Degree distributions shown in Fig.1 include from 4 up to 21 cases (including $k=0$). In particular, the $G(k)$ family includes 6-17 cases, the $G(C)$ includes 5-15 cases, the $G(CB)$ includes 5-19 cases, the $G(CC)$ includes 5-7 cases, and the $G(CE)$ includes 4-21 cases. As it can be observed from Fig.1, frequencies for each null model $G_i$, with $i=1,2,\ldots,30$ (i.e. frequencies included in the *x-z* planes), are arranged into descending order, shaping power-law-alike patterns. Among the five available families, the betweenness $G(CB)$ and eigenvector centrality $G(CE)$ shape the most long-tailed distributions, whereas the family generated by the closeness centrality control-attribute shape the least long-tailed degree distributions.

In order to examine whether degree distributions follow a PL pattern, fittings are applied to all available null models within all five families. Despite the insufficient number of fitting-cases describing very small networks (especially for $n=50$ and 100 nodes), the determination ($R^2$) of the PL-fittings is $\geq 0.874$ in all cases, implying that the degree distribution fit very satisfactorily to PL patterns. In particular, the coefficient of determination for the PL-fittings of the *G*(*k*) family ranges within the interval [0.937, 0.998], of *G*(*C*) within the interval [0.892, 0.997], of *G*(*CB*) within the interval [0.994, 0.999], of *G*(*CC*) within the interval [0.874, 0.996], and of *G*(*CE*) within the interval [0.987, 0.999]. On average, the determination of the PL-fittings are $R_k^2$=0.988, for *G*(*k*), $R_C^2$=0.977, for *G*(*C*), $R_{CB}^2$=0.998, for *G*(*CB*), $R_{CC}^2$=0.937, for *G*(*CC*), and $R_{CE}^2$=0.996, for *G*(*CE*), as it is shown in the 95% confidence intervals (CIs) of Fig.2b.

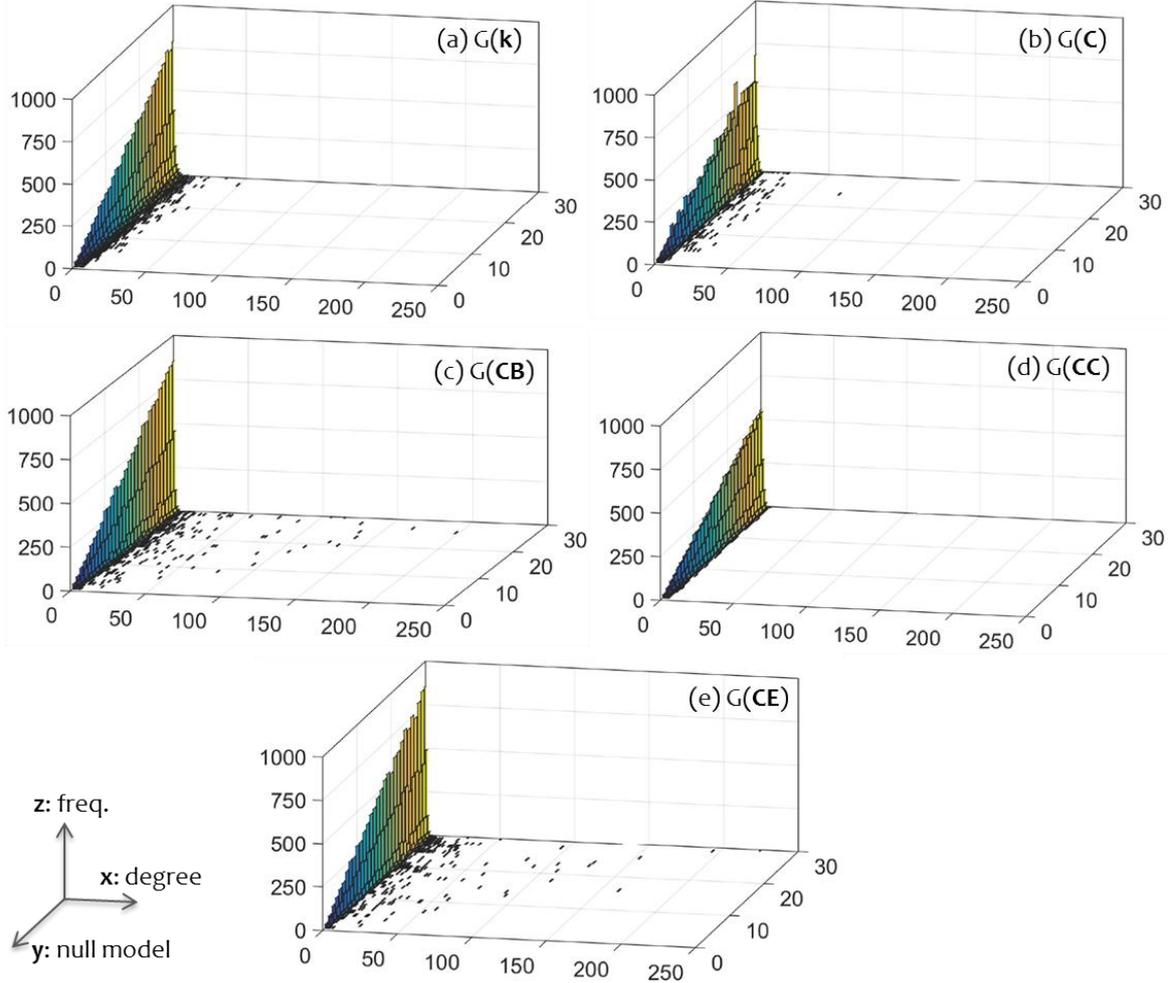

**Fig.1.** Three dimension (3d) bar-charts illustrating the degree distributions of the family of null models: (a) *G*(*k*), which is generated under the control-attribute of degree (*k*), (b) *G*(*C*), which is generated under the control-attribute of clustering coefficient (*C*), (c) *G*(*CB*), which is generated under the control-attribute of betweenness centrality (*CB*), (d) *G*(*CC*), which is generated under the control-attribute of closeness centrality (*CC*), and (e) *G*(*CE*), which is generated under the control-attribute of eigenvector centrality (*CE*). The *x*-axis represents the degree classes, the *y*-axis the null models' ranking (in ascending order according to the number of nodes included in each null model), and the *z*-axis the node-frequencies.

According to Fig.2b, the average determination (<$R^2$>) of PL-fittings are sufficiently high (>0.92) for all the available null-model families, providing strong evidence that the degree distributions follow a PL-pattern and thus that all the available null models are ruled by the SF property. This becomes more evident in Fig.2a, where it can be observed that the average *γ* (gamma) exponents of the PL-fittings are close to the typical interval 2<*γ*<3, which describes real-world networks with the SF property [1]. Especially for the cases of betweenness *G*(*CB*) and eigenvector centrality *G*(*CB*), their total CIs range within this typical interval, implying a perfect compatibility with the empirical observations of the SF real-world networks. Overall, the analysis of the degree distributions shows that the available null models have the SF property and consequently it can be claimed that the GPA process produces SF model for all the examined control-attributes (*k*, *C*, *CB*, *CC*, *CE*), verifying the first research hypothesis.

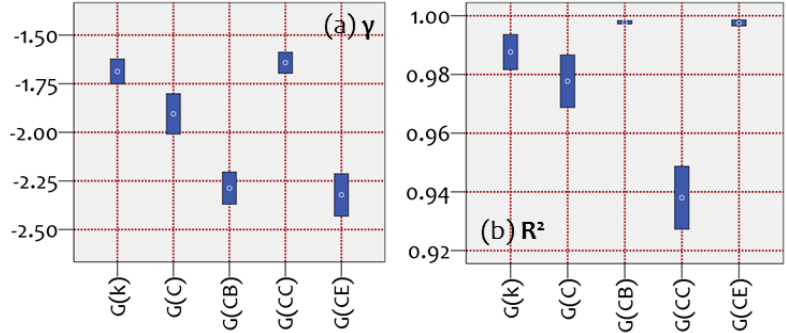

**Fig.2.** 95% confidence intervals (CIs) of the average (a) *poewer-law* (*PL*) *exponent* and (b) *coefficient of determination of the PL-fittings*, computed within each family of networks *G*(*k*), *G*(*C*), *G*(*CB*), *G*(*CC*), and *G*(*CE*). Measures *k*, *C*, *CB*, *CC*, and *CE* within parentheses express the control-attribute under which the generalized preferential attachment (GPA) process is implemented.

### 3.2. Comparisons of network topologies

This part of analysis compares the topological properties of the available families of null models, aiming to detect possible differences between network topologies. At first, the topological layouts of the null models are visualized using the Force-Atlas embedding, which is available in the open-source software of [16]. An indicative picture of the topologies produced by the GPA process is shown in Fig.3, where the null models with *n*=1000nodes are shown, for each network family. As it can be observed, the topological layouts for each null model differ and they configure a characteristic pattern for each family. In particular, the layout of the (betweenness-controlled) null model *G*(*CB*) shows a distinct mono-centric pattern, similar to a superstar network described by the authors of [13], where there is a dominant hub in the network, a considerable concentration of nodes radially to the hub, and a cluster of isolated nodes with an eccentric location in one quadrant of the network space. On the contrary, the layout of the (closeness-controlled) *G*(*CC*) null model shapes a polycentric pattern, where hubs are considerably distant to each other and the isolated along with the other nodes are scattered throughout the network space into a mesh-alike arrangement. The layouts of the degree-controlled *G*(*k*) and eigenvector-controlled *G*(*CE*) null models show a considerable core concentration, similarly with the case of betweenness, but they both considerably differ from the superstar pattern of *G*(*CB*). In particular, *G*(*CE*) shows a distinct polycentric configuration consisting of many hubs, whereas the *G*(*k*) has considerably fewer hubs than *G*(*CE*) and of smaller connectivity (denoted by node size). Also, the degree-controlled null model *G*(*k*) has, similarly with *G*(*CB*), an eccentric cluster of isolated nodes located in one quadrant of the network space, whereas the eigenvector-controlled *G*(*CE*) null model has its isolated nodes located into a ring arrangement covering all the network space. Overall, in Fig.3, we can observe a variety of topologies ranging from a mesh-alike to a superstar-alike pattern according to the ordering *G*(*CC*), *G*(*C*), *G*(*CE*), *G*(*k*), and *G*(*CB*).

At the next step, a variety of network measures, metrics, and statistics are examined for the available families of null models, in order to detect statistical differences among the topologies of these families. This approach conceptualizes network topology as the composition of the available network metrics, each of which measures a certain aspect of network topology [7,21]. The network measures participating in this analysis are briefly described in Table 1.

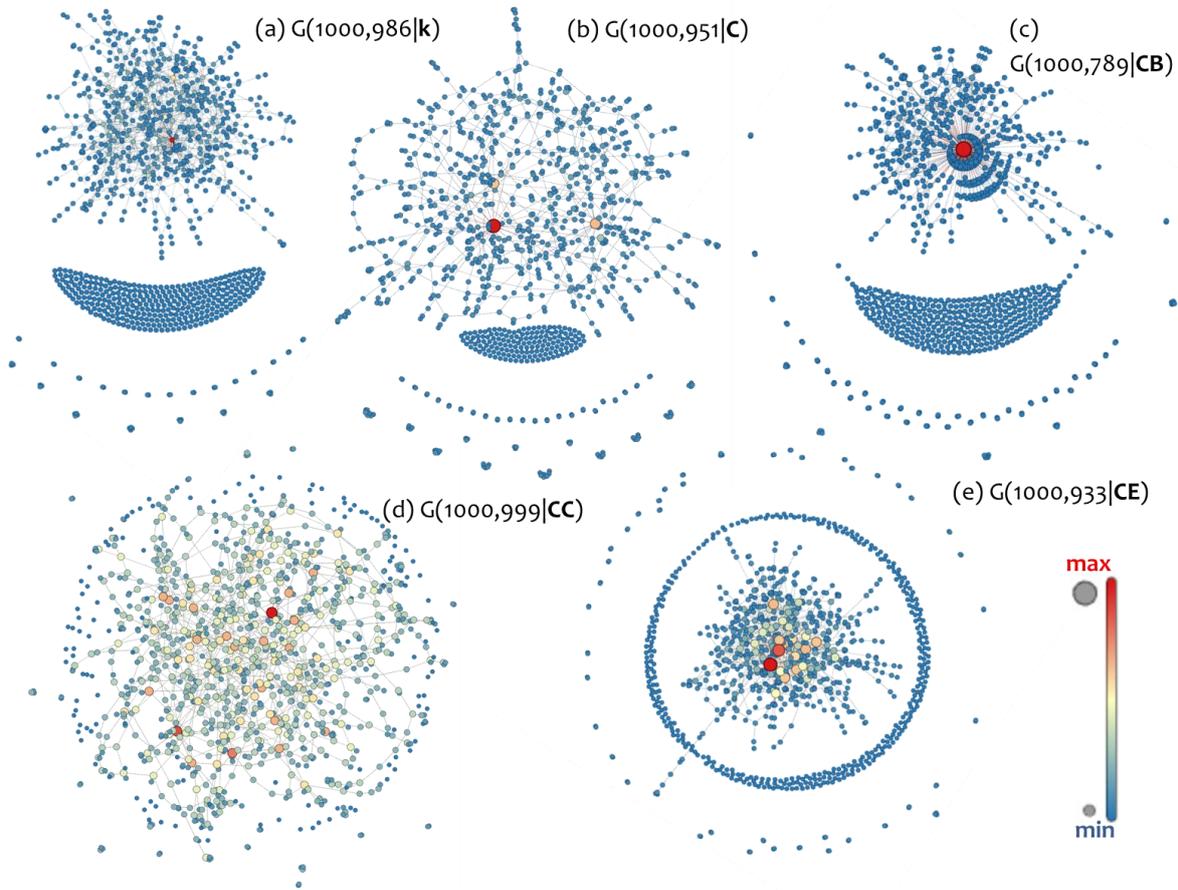

**Fig.3.** Topological layouts of null models $G(n,m|X)$, with $n=1000$ nodes and $m$ edges, which are generated under the control-attribute of (a) degree ($X=k$, $m=986$), (b) clustering coefficient ($X=C$, $m=951$), (c) betweenness centrality ($X=CB$, $m=789$), (d) closeness centrality ($X=CC$, $m=999$), and (e) eigenvector centrality ($X=CE$, $m=933$). Layouts are visualized by using the (force-directed) Force-Atlas embedding, which is available in the open-source software of [16]. Node color (from blue to red) and size (from small to big) are shown proportionally to node degree.

Table 1
Network measures considered in the topological analysis

| Measure | Description | Math Formula | Reference(s) |
|---|---|---|---|
| Network | A graph, expressed as the pair set of nodes $V$ and edges $E$. | $G(V,E)$ | [13] |
| Network edges ($m$) | The number of links included in the network | $m=|E|=\text{card}(E)$ | [13] |
| Diagonal Distance ($dd$) | The average distance of the non-zero elements from the main diagonal of the network's adjacency. | $dd(G) = \dfrac{1}{\sqrt{2} \cdot n^2} \sum_{(i,j) \in E} |i-j|$ | [13] |
| Network diameter $d(G)$ | The longest path in the network. | $dG = \max\{d(i,j) \mid i,j \in V\}$ | [13] |

| Measure | Description | Math Formula | Reference(s) |
|---|---|---|---|
| Node Degree ($k$) | Number of edges being adjacent to a node. | $k_i = k(i) = \sum_{j \in V} \delta_{ij}$, where $\delta_{ij} = \begin{cases} 1, & \text{if } e_{ij} \in E \\ 0, & \text{otherwise} \end{cases}$ | [1,13] |
| Maximum degree ($k_{max}$) | The maximum degree of the network nodes. | $k_{max} = \max\{k(i) \in V \mid i = 1, 2, ..., n\}$ | [13] |
| Isolated nodes ($k_o$) | The number of unconnected ($k=0$) nodes in the network. | $k_o = card\{k(i) = 0 \mid i = 1, 2, ..., n\}$ | [13] |
| UDV | *Unique degree values*: the number of distinct degrees considered for computing the degree distribution of a network. | n/a | In this paper |
| Hubs | The number of network nodes with degree within the last tenth of the degree-range. | $Hubs = \{i \in V \mid k(i) \geq k_{max} - (\frac{k_{max} - k_{min}}{10})\}$ | In this paper |
| Average Path Length $\langle l \rangle$ | Average network shortest path lengths $d(i,j)$. | $\langle l \rangle = \frac{\sum_{v \in V} d(v_i, v_j)}{n \cdot (n-1)}$ | [8,13] |
| COM | Number of connected components in the network. | | [7,14] |
| Assortativity ($r$) | A measure of nodes' preference to attach to other similar nodes, where $e_{jk}$ is the joint probability distribution of the remaining degrees of two nodes at either end of a randomly chosen end. | $r = \frac{1}{\sigma_q^2} \sum_{jk} jk(e_{jk} - q_j q_k)$, where $\sum_{jk} e_{jk} = 1$ and $\sum_j e_{jk} = q_k$ | [7,20] |
| Local Clustering Coefficient ($C(i)$) | The number of a node's connected neighbors $E(i)$, divided by the number of the total triplets $k_i(k_i-1)$ shaped by the node. | $C(i) = \frac{E(i)}{k_i \cdot (k_i - 1)}$ | [8] |
| Modularity ($Q$) | Objective function measuring the potential of a network to be subdivided into communities, where $g_i$ is the community of node $i$, $[A_{ij} - P_{ij}]$ is the actual minus the expected number of edges falling between a particular pair of nodes. | $Q = \frac{\sum_{i,j}[A_{ij} - P_{ij}] \cdot \delta(g_i, g_j)}{2m}$, where $\delta_{ij} = \begin{cases} 1, & \text{if } g_i = g_j \\ 0, & \text{otherwise} \end{cases}$ | [4] |
| $\omega$-index | Index detecting whether a network has the small-world property, or lattice-like, or random-like characteristics. | $\omega = \frac{\langle l \rangle_{rand}}{\langle l \rangle} - \frac{\langle C \rangle}{\langle C \rangle_{latt}}$ | [22] |
| COI | *City organization index*: small values ($\approx 0$) express that the network is described by well-organized pattern. Values close to one ($\approx 1$) express deficiency in organization and planning. | $r_n = \frac{n(1) + n(3)}{\sum_{k \neq 2} n(k)}$ | [23] |

For the comparison of network topologies between null-model families, a statistical-inference analysis is applied to the set of measures shown in Table 1. The results of the analysis are shown in Fig.4, where 95% confidence intervals (CIs) [24] of the mean-values are displayed per measure and per family. Non-overlaid intervals, in Fig.4, imply statistical differences between compared cases, whereas overlaid intervals do not. Generally, as it can be observed, in plenty of cases the CIs do not overlay, implying that the network topologies of the null-model families differ in many aspects, at least pair-wisely. The only wide exception to this observation concerns the number of links, where its average values are not statistically different for any pair of families. All the other cases provide insights supporting the statistical difference of the examined null-model families. In particular, Fig.4b shows the CIs of diagonal distance (dd), a spectral measure proposed by the author of [7] for the detection of topological differences between networks (and particularly of SF), where it can be observed that the betweenness-controlled family $G(CB)$ has the most concentrated sparsity pattern to the main diagonal of the adjacency, whereas the closeness-controled family $G(CC)$ has the most scattered. Provided that node-ages are known due to algorithm-generated process of GPA, statistical differences among these cases of dd(G) imply a significant possibility the overall topologies of the null-model families to different according to the CIs' difference [7]. Next, in Fig.4c, the CIs of network diameter imply that the betweenness-controlled family $G(CB)$ has, on average, the shortest diameter (however it is not statistically different with the $G(CE)$), whereas the closeness-controlled family $G(CC)$ has the largest network diameter (not statistically different with the $G(C)$). In Fig.4d, it can be observed that the closeness-controlled family $G(CC)$ has the highest average degree, whereas the betweenness-controlled $G(CB)$ and eigenvector-controlled $G(CE)$ families have the smallest. This complies with the observation in the topological layouts (Fig.3), according to which the topologies of $G(CB)$ and $G(CE)$ are more hub-and-spoke-alike, in contrast with the more mesh-alike topology of $G(CC)$. This picture becomes more complete with Fig.4e, according to which the maximum degree of the betweenness-controlled family $G(CB)$ is statistically greater than all other cases. This supports the superstar-alike observation made for the topology of $G(CB)$. The next Fig.4f shows that the clustering-controlled $G(C)$ and closeness-controled $G(CC)$ families have the fewest number of isolated nodes in their networks, supporting the observation made in Fig.3 about their mesh-alike topologies. Next, Fig.4g shows that the closeness-controled family $G(CC)$ has less unique degree-values than the other cases, implying that it has the least long-tailed degree distribution and thus higher probability to include more hubs than the other cases. This interpretation is supported by Fig.4h, which also complies with the observation in the topological layouts (Fig.3) about the mesh-alike topology of $G(CC)$. A similar to the network diameter picture is also shaped in Fig.4i, which shows the CIs of the average path length. This picture complies with the mesh-alike to superstar-alike ordering of topologies made in the previous section. Next, Fig.4j shows the number of components in the network, which is similar to the case of isolated nodes in Fig.4f. In Fig.4k, the degree-controlled family $G(k)$ is the most assortative family than the others, implying a good tendency of nodes to attach with similar ones. Next, Fig.4l shows the CIs of the average clustering and clustering coefficients. Impressively enough, although clustering was used as control-attribute in the GPA process, the betweenness-controlled family $G(CB)$ is the highest clustered family among the others. In Fig.4m, the clustering-controlled $G(C)$ and closeness-controlled $G(CC)$ families have the best tendency to be divided into communities, which is something obviously related to their better mesh-alike topology profiles (Fig.3).

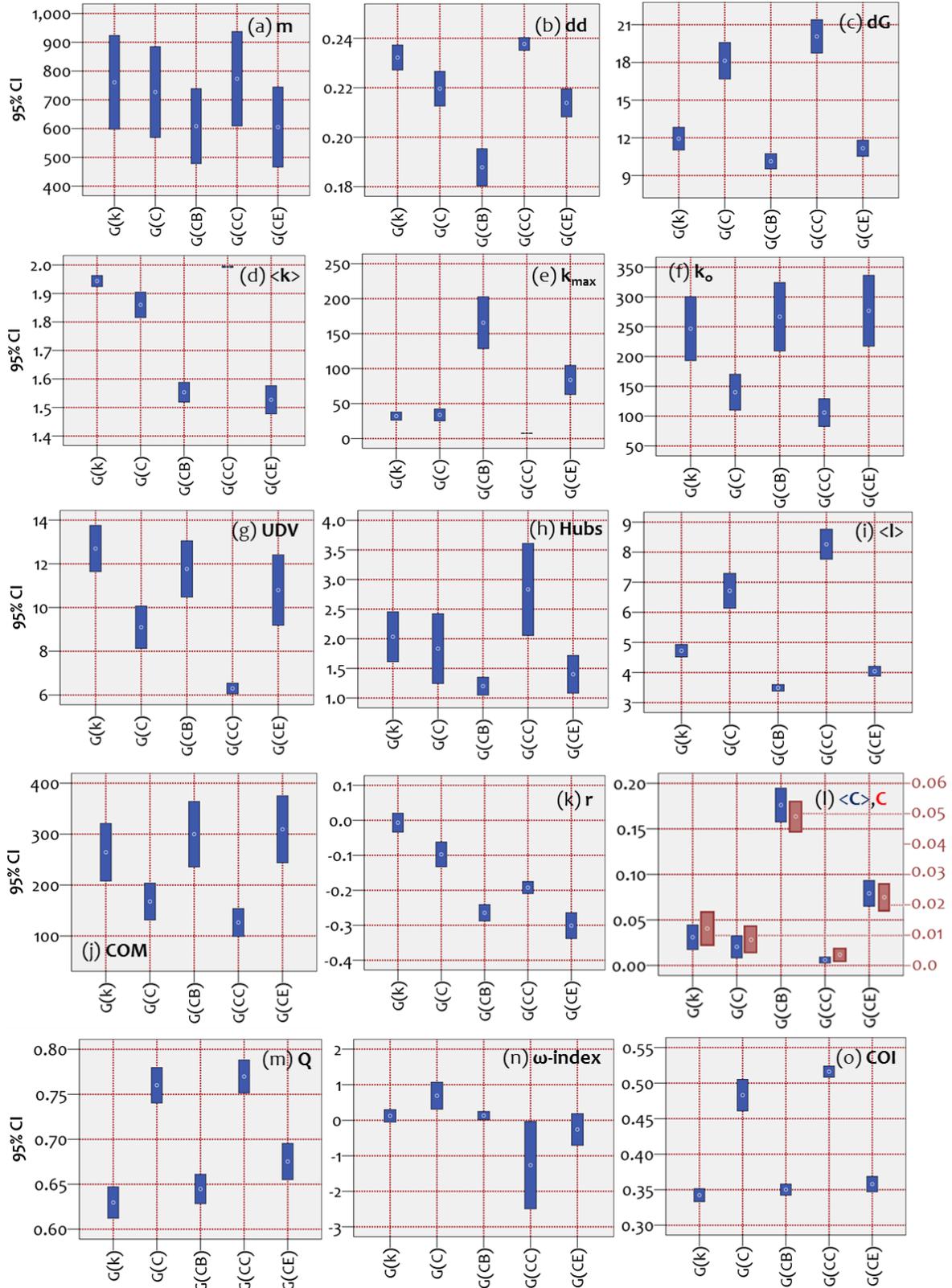

**Fig.4.** 95% confidence intervals (CIs) of the average (a) *number of links*, (b) *diagonal distance* (see [7]), (c) *network diameter*, (d) *average degree*, (e) *max degree*, (f) *number of isolated nodes*, (g) *unique degree values*, (h) *number of hubs*, (i) *average path length*, (j) *number of connected components*, (k) *network assortativity*, (l) *average and global clustering coefficient*, (m) *modularity*, (n) *ω-index*, proposed by [22], and (o) *city organization index*, proposed by [23]. All CIs were computed within each family of networks *G(k)*, *G(C)*, *G(CB)*, *G(CC)*, and *G(CE)*.

Measures *k*, *C*, *CB*, *CC*, and *CE* within parentheses express the control-attribute under which the generalized preferential attachment (GPA) process is implemented.

In Fig.4n, a better performance is shaped for those CIs being closer to the zero-line, because this condition implies a small-world-alike (SW-alike) topology (for SW networks see [25]). Under this condition, the betweenness-controlled *G*(*CB*) and degree-controlled *G*(*k*) families have closer to SW-alike topologies. Finally, Fig.4o shows the CIs of the city organizations index, which was borrowed from the study of spatial networks [23] to provide insights about the network topology. Based on the COI, the degree-controlled *G*(*k*), betweenness-controlled *G*(*CB*), and eigenvector-controlled *G*(*CE*) families are described by the most well-organized patterns in their network topologies.

Overall, this analysis showed that plenty of aspects of the examined topological attributes differ (either pair-wisely or in total) between the null-model families. This verifies the second research question interpreting that the network topologies produced by the generalized PA (GPA) process under different control-attributes are different.

### 3.3. Extracting the optimum network topology

The final part examines which among the available families of null models configures the best network topology. The answer of this question is based on the CIs shown in Fig.2 and Fig.4. At first, Table 2 is constructed, summarizing the cases where minimum or maximum CIs are observed in Fig.2 and Fig.4. The summary of Table 2 shows that the betweenness-controlled family *G*(*CB*) appears 8 times optimum performance, being followed by the eigenvector-controlled *G*(*CE*), closeness-controlled *G*(*CC*), and degree-controlled *G*(*k*) and clustering-controlled *G*(*C*) families.

**Table 2**
Summary of measures with minimum or maximum CIs[a]

| Measure | Optimum condition[b] | Null-model family | | | | |
|---|---|---|---|---|---|---|
| | | *G*(*k*) | *G*(*C*) | *G*(*CB*) | *G*(*CC*) | *G*(*CE*) |
| dd | n/a | | | min | | |
| <*k*> | max | | | min | max[*] | min |
| *r* | min | max | | min[*] | | min[*] |
| $k_{max}$ | max | | | max[*] | min | |
| $k_o$ | min | | min[*] | | min[*] | |
| UDV | max | | | | min | |
| γ | 2 < γ < 3 | | | min[*] | | min[*] |
| $R^2$ | →1 | | | | min | |
| ω-index | →0 | [*] | max | [*] | min | min |
| COI | →0 | min[*] | | min[*] | max | min[*] |
| d*G* | min | | | min[*] | | min[*] |
| *Q* | max | min[*] | | min[*] | | min[*] |
| COM | min | | min[*] | | min[*] | |
| <*C*> | max | | | max[*] | min | |
| *C* | max | | min | max[*] | min | |
| <*l*> | min | | | min[*] | max | |
| Min | | 2 | 2 | 8 | 8 | 7 |
| Max | | 1 | 2 | 3 | 3 | 0 |
| Optimums[*] | | 2 | 2 | 8 | 4 | 5 |

a. Based on the analysis shown in Fig.2 and Fig.4.
b. Defined by the physical meaning of each measure, based on relevant literature (see Table 1)

In order to count all possible pair-wise statistical differences observed in the analysis of Fig.2 and Fig.4, a comparative directed graph is constructed. In this graph, each null-model family is assigned to a node, where a directed link between nodes $i,j$ is created when node $i$ outperforms (has more optimum topological performance than) node $j$, according to the relation:

$$i \rightarrow j \equiv e_{ij} \in E \,|\, \text{CI}(i) \overset{optimum}{>} \text{CI}(j) \qquad (3).$$

Based on this criterion, the graph of Fig.5 is constructed, where the weighted out-degree indicates the state of outperformance in terms of the aggregated topological attributes of Fig.2 and Fig.4. Within this context, the better topological profile is shaped, in descending order, by the null model families $G(CB)$, $G(CE)$, $G(C)$, $G(k)$, and $G(CC)$.

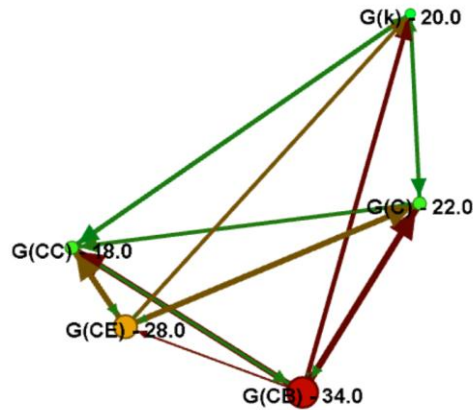

**Fig.5.** Comparative directed graph of optimum performance according to the CIs observed in Fig.2 and Fig.4. In this graph, each null-model family ($G(k)$, $G(C)$, $G(CB)$, $G(CC)$, $G(CE)$) is assigned to a node, where a directed link between nodes $i,j$ is created when node $i$ outperforms (has more optimum performance than) node $j$. Nodes are colored and sized proportional to the weighted out-degree, where higher values indicate more optimum topological performance.

Overall, according to the previous analysis, the null models generated by the GPA process under the control-attribute of betweenness centrality were found to have the most optimum topology, in terms of the topological aspects considered in the analysis (Table 1). This finding verifies the result of [25], who observed that the PA process based on node-betweenness suggests a better indicator of social attractiveness, along with the observation of the authors of [20], who noted that superstar SF networks are of better topology of the BA model.

## 5. Conclusions

This paper expanded the degree-based consideration of the growth process of preferential attachment (PA), by considering five different connectivity criteria (node degree, clustering coefficient, betweenness centrality, closeness centrality, and eigenvector centrality) as attractors (control-attributes) for the development of new links in the network. The analysis was coordinated by three research questions, the first was whether the PA process can generate scale-free (SF) networks for every control-attribute, the second was whether the network topologies produced by the generalized PA (GPA) process are different, and, the third was which control-attribute is capable producing a better network topology. Based on statistical inference applied to various measures of network topology the analysis showed that all the available control attributes are capable generating SF networks. However, in the majority of cases the examined topological attributes were statistically different (at least pair-wisely) implying that the generalized PA (GPA) process produces networks of different topologies, under different control-attributes. Finally, betweenness centrality was found to be the control-attribute generating networks of better topology. Overall, by answering these questions, this paper introduces a multi-dimensional conceptualization of the preferential attachment growth process, which can motivate further research and can provide new tools for the modeling and interpretation of real-world networks that currently cannot be fully explained by the degree-driven BA model.

# Appendix

**Table 2**
The null models generated by the generalized preferential attachment growth process and participated in the analysis

| | | Null-model family | | | | |
|---|---|---|---|---|---|---|
| | $i=$ | $G_i(k)$ (degree-controlled) | $G_i(C)$ (clustering-controlled) | $G_i(CB)$ (betweenness-controlled) | $G_i(CC)$ (closeness-controlled) | $G_i(CE)$ (eigenvector-controlled) |
| Null-model ranking | 1 | $G_1(50,44)$ | $G_1(50,45)$ | $G_1(50,32)$ | $G_1(50,50)$ | $G_1(50,38)$ |
| | 2 | $G_2(100,94)$ | $G_2(100,97)$ | $G_2(100,73)$ | $G_2(100,99)$ | $G_2(100,74)$ |
| | 3 | $G_3(150,140)$ | $G_3(150,147)$ | $G_3(150,106)$ | $G_3(150,149)$ | $G_3(150,107)$ |
| | 4 | $G_4(200,184)$ | $G_4(200,195)$ | $G_4(200,163)$ | $G_4(200,199)$ | $G_4(200,146)$ |
| | 5 | $G_5(250,245)$ | $G_5(250,191)$ | $G_5(250,192)$ | $G_5(250,251)$ | $G_5(250,191)$ |
| | 6 | $G_6(300,278)$ | $G_6(300,278)$ | $G_6(300,230)$ | $G_6(300,297)$ | $G_6(300,223)$ |
| | 7 | $G_7(350,343)$ | $G_7(350,276)$ | $G_7(350,294)$ | $G_7(350,348)$ | $G_7(350,261)$ |
| | 8 | $G_8(400,384)$ | $G_8(400,374)$ | $G_8(400,313)$ | $G_8(400,397)$ | $G_8(400,278)$ |
| | 9 | $G_9(450,440)$ | $G_9(450,376)$ | $G_9(450,349)$ | $G_9(450,449)$ | $G_9(450,340)$ |
| | 10 | $G_{10}(500,492)$ | $G_{10}(500,464)$ | $G_{10}(500,365)$ | $G_{10}(500,499)$ | $G_{10}(500,362)$ |
| | 11 | $G_{11}(550,539)$ | $G_{11}(550,522)$ | $G_{11}(550,471)$ | $G_{11}(550,548)$ | $G_{11}(550,411)$ |
| | 12 | $G_{12}(600,588)$ | $G_{12}(600,595)$ | $G_{12}(600,404)$ | $G_{12}(600,598)$ | $G_{12}(600,441)$ |
| | 13 | $G_{13}(650,636)$ | $G_{13}(650,593)$ | $G_{13}(650,510)$ | $G_{13}(650,645)$ | $G_{13}(650,461)$ |
| | 14 | $G_{14}(700,689)$ | $G_{14}(700,696)$ | $G_{14}(700,582)$ | $G_{14}(700,695)$ | $G_{14}(700,514)$ |
| | 15 | $G_{15}(750,732)$ | $G_{15}(750,704)$ | $G_{15}(750,596)$ | $G_{15}(750,747)$ | $G_{15}(750,553)$ |
| | 16 | $G_{16}(800,788)$ | $G_{16}(800,735)$ | $G_{16}(800,643)$ | $G_{16}(800,794)$ | $G_{16}(800,575)$ |
| | 17 | $G_{17}(850,828)$ | $G_{17}(850,825)$ | $G_{17}(850,669)$ | $G_{17}(850,851)$ | $G_{17}(850,606)$ |
| | 18 | $G_{18}(900,890)$ | $G_{18}(900,809)$ | $G_{18}(900,738)$ | $G_{18}(900,899)$ | $G_{18}(900,643)$ |
| | 19 | $G_{19}(950,935)$ | $G_{19}(950,903)$ | $G_{19}(950,742)$ | $G_{19}(950,948)$ | $G_{19}(950,702)$ |
| | 20 | $G_{20}(1000,986)$ | $G_{20}(1000,951)$ | $G_{20}(1000,789)$ | $G_{20}(1000,999)$ | $G_{20}(1000,933)$ |
| | 21 | $G_{21}(1050,1035)$ | $G_{21}(1050,1011)$ | $G_{21}(1050,791)$ | $G_{21}(1050,1048)$ | $G_{21}(1050,775)$ |
| | 22 | $G_{22}(1100,1088)$ | $G_{22}(1100,1006)$ | $G_{22}(1100,859)$ | $G_{22}(1100,1099)$ | $G_{22}(1100,886)$ |
| | 23 | $G_{23}(1150,1132)$ | $G_{23}(1150,1073)$ | $G_{23}(1150,885)$ | $G_{23}(1150,1149)$ | $G_{23}(1150,844)$ |
| | 24 | $G_{24}(1200,1191)$ | $G_{24}(1200,993)$ | $G_{24}(1200,977)$ | $G_{24}(1200,1196)$ | $G_{24}(1200,873)$ |
| | 25 | $G_{25}(1250,1229)$ | $G_{25}(1250,1248)$ | $G_{25}(1250,993)$ | $G_{25}(1250,1248)$ | $G_{25}(1250,1170)$ |
| | 26 | $G_{26}(1300,1281)$ | $G_{26}(1300,1226)$ | $G_{26}(1300,977)$ | $G_{26}(1300,1296)$ | $G_{26}(1300,1000)$ |
| | 27 | $G_{27}(1350,1325)$ | $G_{27}(1350,1323)$ | $G_{27}(1350,1045)$ | $G_{27}(1350,1349)$ | $G_{27}(1350,1246)$ |
| | 28 | $G_{28}(1400,1373)$ | $G_{28}(1400,1368)$ | $G_{28}(1400,1061)$ | $G_{28}(1400,1400)$ | $G_{28}(1400,1237)$ |
| | 29 | $G_{29}(1450,1437)$ | $G_{29}(1450,1450)$ | $G_{29}(1450,1186)$ | $G_{29}(1450,1448)$ | $G_{29}(1450,1131)$ |
| | 30 | $G_{30}(1500,1475)$ | $G_{30}(1500,1332)$ | $G_{30}(1500,1219)$ | $G_{30}(1500,1497)$ | $G_{30}(1500,1135)$ |